\documentclass[aps,pre,preprint,groupedaddress]{revtex4}

\usepackage{graphicx}
\newcommand{\pfrac}[2]{\frac{\partial #1}{\partial #2}}

\newcommand{\vrp}{\vec{r'}}
\newcommand{\vq}{\vec{q}}

\newcommand{\wrho}{\widehat{\rho}}

\begin{document}


\title{Structure factor of
  polymers interacting via a short range repulsive potential: application to
  hairy wormlike micelles}


\author{Gladys Massiera}
\altaffiliation{present address: Chemical and Biomolecular
Engineering Department, University of Pennsylvania, Towne 311A,
Philadelphia, PA 19104, USA.}
 \affiliation{Groupe de Dynamique des Phases Condens\'{e}es
(CNRS-UM2 n$^o$5581), CC26, Universit\'{e} Montpellier 2, 34095
Montpellier Cedex 5, France.}

\author{Estelle Pitard}
\affiliation{Laboratoire de Physique Math\'{e}matique et
Th\'{e}orique (CNRS-UM2 n$^o$5825), CC50, Universit\'{e}
Montpellier 2, 34095 Montpellier Cedex 5, France.}

\author{Laurence Ramos}
\affiliation{Groupe de Dynamique des Phases Condens\'{e}es
(CNRS-UM2 n$^o$5581), CC26, Universit\'{e} Montpellier 2, 34095
Montpellier Cedex 5, France.}

\author{Christian Ligoure}
\affiliation{Groupe de Dynamique des Phases Condens\'{e}es
(CNRS-UM2 n$^o$5581), CC26, Universit\'{e} Montpellier 2, 34095
Montpellier Cedex 5, France.}


\date{\today}

\begin{abstract}

We use the  Random Phase Approximation (RPA) to compute the
structure factor, $S(q)$, of a solution of chains interacting
through a soft and short range repulsive potential $V$. Above a
threshold polymer concentration, whose magnitude is essentially
controlled by the range of the potential, $S(q)$ exhibits a peak
whose position depends on the concentration. We take advantage of
the close analogy between polymers and wormlike micelles and apply
our model, using a Gaussian function for $V$, to quantitatively
analyze experimental small angle neutron scattering profiles of
semi-dilute solutions of hairy wormlike micelles. These samples,
which consist in surfactant self-assembled flexible cylinders
decorated by amphiphilic copolymer, provide indeed an appropriate
experimental model system to study the structure of sterically
interacting polymer solutions.

\end{abstract}

\pacs{61.25.Hq, 83.80.Qr, 61.12.-q, 05.20.-y}

\maketitle



\section{INTRODUCTION}

Isotropic solutions of uni-dimensional objects like polymers do
not generally show a correlation peak in their structure factor
except for very concentrated systems and melts, or charged
systems. For polyelectrolytes, the peak originates from strong
electrostatic interchains interactions, whereas for neutral
polymers the peak is observed at very large scattering vector, on
the order of the inverse of the monomer length, and is the
signature of  a liquid-like order at the monomer length scale, as
in simple liquids. Recently we have reported on a new type of
living polymer system which also exhibits a structural peak in the
scattering function, but with a totally different physical origin
\cite{massiera-langmuir}. The experimental system is a semi-dilute
solution of hairy wormlike micelles, obtained by adding small
amounts of amphiphilic copolymer to a solution of surfactant
micelles. The correlation peak, observed even at low concentration
of micelles but still in the semi-dilute regime, originates from
steric repulsion between the  micelles, induced by the copolymer
layer which covers them. Despite its short range nature, on the
order of the copolymer layer thickness, this interaction is
sufficient to generate a correlation peak, as demonstrated by
small angle neutron scattering (SANS) experiments. Because
surfactant self-assemblies scatter light and neutron much more
strongly than polymers, surfactant wormlike micelles have appeared
as a convenient model system for the study of the structure of
polymer solutions \cite{walker2001}. In particular, charged
wormlike micelles have been extensively studied as an alternative
system to conventional polyelectrolyte solutions
\cite{magid1998,schmitt1995,cannavacciuolo2002-2}. Similarly,
hairy wormlike micelles may be a suitable model system to
investigate steric interactions in semi-dilute solutions of
polymers.

The Random Phase Approximation (RPA) represents a powerful
theoretical tool to predict the structure factor of polymeric
systems, provided concentration fluctuations are weak. Introduced
a long time ago in the context of simple liquids
\cite{AndChandler,Hansen}, it has been reformulated explicitly by
Edwards in the framework of polymer theory \cite{edwards1966}, and
been applied since to numerous polymer systems, for instance to
polyelectrolytes solutions
\cite{barrat1996,borue1988,joanny1990,castelnovo2001} or to
microphase separation in block copolymers melts
\cite{leibler1980}. For concentrated polymer solutions,
concentration fluctuations are weak, allowing a perturbative
approach for the calculation of correlation functions, starting
from the mean field Hamiltonian of the system. A perturbative
calculation around the homogeneous equilibrium values of the
monomer densities allows in particular the computation of the
structure factor. In this paper, we add a Gaussian repulsive
potential to the classical excluded volume interactions between
monomers, using a RPA description of polymer solutions  (using
Edward's formalism) and calculate the structure factor $S(q)$. We
show that $S(q)$ may exhibit a broad correlation peak, whose
existence and position as a function of polymer concentration,
range and magnitude of the Gaussian potential are discussed. The
behavior of the theoretical $S(q)$ appears in excellent
concordance with the experimental SANS scattering profiles of
hairy wormlike micelles. Moreover, we use our model to fit the
experimental peak position as a function of the micellar
concentration and derive a measurement of the thickness $h$ of the
polymer layer covering the micelle. The numerical values of $h$
are found in good agreement with simple theoretical expectations
and with other experimental determinations \cite{MassieraLiqMatt}.

The paper is organized as follows: Section~\ref{SEC:exp} describes
the experimental system and recalls the main experimental results
reported in Ref.\cite{massiera-langmuir} concerning the scattering
patterns of semi-dilute solutions of hairy wormlike micelles. In
Sec.~\ref{SEC:Theo}, we describe our model based on the RPA
technique and report technical details of the calculations in the
appendix. The model, which allows one to predict the structure
factor of semi-dilute polymer solutions which interact via a
short-range repulsive potential, is applied to our specific
experimental system. Finally, in Sec.~\ref{SEC:discuss} we compare
the model to experiments and derive a quantitative evaluation for
the thickness of the polymeric layer. We conclude in
Sec.~\ref{SEC:concl} by some remarks concerning the generality of
our approach and the validity of the RPA in the framework of
semi-dilute solutions of wormlike micelles.


\section{Experiments}
\label{SEC:exp}

\subsection{Experimental system}

Hairy polymers are obtained by adding to solutions of wormlike
micelles small amounts of amphiphilic copolymer, whose hydrophobic
part adsorbs onto the fluid surfactant cylinders and whose
hydrophilic tail remains swollen in water and decorates the
micelles. The surfactant micelles are formed by diluting in brine
([NaCl]=$0.5 \, \rm{M}$) a mixture of cetylpyridinium chloride
(CpCl) and sodium salicylate (NaSal) at a fixed molar ratio
$\rm{[NaSal]/[CpCl]}=0.5$ \cite{Rehage-Hoffmann}. We use
commercially available triblock copolymers (Synperonic F108 and
F68, by Serva, used as received) and a diblock copolymer (PC18,
synthesized in our laboratory \cite{hartmann1999}). The Synperonic
F108 (F68) consists in two identical hydrophilic polyoxyethylene
(POE) blocks of 127 (76) monomers each, symmetrically bounded to a
central shorter hydrophobic block of polyoxypropylene (PPO) of 48
(29) monomers. The polymer PC18 consists in a $C_{18}$ alkyl chain
as hydrophobic part, bounded by an uretane group to a POE block of
113 monomers. The radii of gyration of the hydrophilic blocks are
$18.6$, $25.8$, and $24\ $\AA\ for F68, F108 and PC18
respectively. We have shown in Ref.\cite{massiera-langmuir} that
the cylindrical structure of the micelles is maintained upon
copolymer addition, with a constant radius of their hydrophobic
core $r_C \approx 21\ \rm{\AA}$. We define $\phi$ as the
surfactant volume fraction and $\alpha$ as the PEO-block to
surfactant molar ratio. In our experiments, we vary $\phi$ and
$\alpha$ between $2.8$ and $40 \%$ and $0$ and $4.2 \%$
respectively. The parameter $\alpha$ controls the density of the
polymeric layer. The crossover $\alpha^*$ between the mushroom
regime and the brush regime \cite{DeGennes1980} being estimated to
$3 \%$, $1.5\ \%$ and $1.8\ \%$ for F68, F108 and PC18
respectively, both regimes are probed in our experiments. On the
other hand, in the range of surfactant concentration investigated,
the micellar solutions are in the semi-dilute regime.

SANS experiments are performed on the spectrometer PACE at the
Laboratoire L\'{e}on Brillouin (Saclay, France) and on the D11
beam line at the Institut Laue Langevin (Grenoble, France). We use
deuterated water, all other components being hydrogenated. Neutron
experiments are thus sensitive to the contrast between the
hydrophobic core of the micelles and the aqueous solvent
($\rm{D_{2}0}$). In particular, because the hydrophilic POE blocks
of the copolymer are always highly swollen in $\rm{D_{2}0}$, the
contribution of the copolymer layer to the scattered intensity is
negligible compared to the contribution of the hydrophobic core
and thus the copolymer layer covering the micelles is not directly
probed. In all the experiments, the temperature is fixed at $30 \
^{\rm{o}}\rm{C}$.

\subsection{Correlation peak}

In this section, we recall the main experimental results
previously reported by us in Ref.\cite{massiera-langmuir}. Figure
\ref{exp-phivar}a shows the variation of the scattering profile
for samples with constant copolymer density ($\alpha=1 \, \%$) but
with various surfactant volume fractions $\phi$. The scattering
profile is monotonically decreasing at low $\phi$ and above a
threshold surfactant volume fraction $\phi_{c}$, a correlation
peak is observed at a finite wavevector. The intensity of the peak
increases with $\phi$ and its position $q^*$ is reported to higher
wavevector when $\phi$ increases. Figure \ref{exp-alphavar}a shows
the variation of the scattering profile at a fixed surfactant
volume fraction $\phi=9 \, \%$ when the copolymer density is
increased. A peak emerges above a threshold copolymer molar ratio
and becomes more and more pronounced and narrow as $\alpha$
increases. Moreover, the peak position varies only weakly with the
copolymer to surfactant ratio. We note that these features are
obtained for the three types of copolymer used.

Because of the high ionic strength ($0.5 \, \rm{M}$),
electrostatic interactions are screened and are thus not relevant
in our experiments. The correlation peak observed experimentally,
whose intensity increases with the copolymer over surfactant
ratio, originates therefore from the copolymer layer adsorbed onto
the micelles. This layer creates a steric short-range repulsion
between the micelles, with a range on the order of the copolymer
layer thickness. To analyze more quantitatively the scattering
profiles, we model the soft short-range copolymer-induced
repulsion and use the RPA technique to compute the structure
factor of a solution of polymeric chains interacting via a
short-range repulsion.

\section{Theoretical analysis}
\label{SEC:Theo}

\subsection{RPA Model}

In a continuous approach, the Hamiltonian ${\LARGE
\mathcal{H}}({\vec{r}(s)})$ of a linear chain of $N$ statistical
units in a given configuration $\vec{r}(s)$ reads, in $k_BT$ units
:
\begin{eqnarray}
{\LARGE \mathcal{H}}({\vec{r}(s)}) &=& \frac{3}{2a^2}\
\int_{0}^{N} ds \left( \pfrac{\vec{r}}{s} \right)^2 +  \nonumber
\\ & & + \frac{1}{2} \int_0^N\ \int_0^N\ ds\ ds'\
\mathbf{V}\left(\vec{r}(s) -
\vec{r}(s')\right)\label{free-energy-RPA}
\end{eqnarray}
where $a$ is the monomer size, $\vec{r}(s)$ is the position of the
$s^{th}$-monomer, and $\mathbf{V}(\vec{r} - \vec{r'})$ is the
monomer-monomer interaction potential. The first term of the right
end side of Eq.\ref{free-energy-RPA} represents the entropic
contribution of the chain configurations (the ``entropic
elasticity'') and the second term describes the two-body
interaction. From this Hamiltonian, we calculate the partition
function of the system, and we evaluate the monomer density
auto-correlation function, which is directly proportional to the
structure factor $S(q)$. The details of these calculations are
given in the appendix and can be generalized to a system of $M$
independent chains containing each $N$ monomers
\cite{doi-edwards}. The result is a classical general expression
for the structure factor for any given microscopic potential
$\mathbf{V}(\vec{r})$ between monomers:

\begin{equation}
S^{-1}(\vec{q}) = S_{0}^{-1}(\vec{q}) +
\widetilde{\mathbf{V}}(\vec{q}), \label{RPAeq}
\end{equation}

\noindent where $\widetilde{\mathbf{V}}(\vec{q})$ is the Fourier
transform of $\mathbf{V}(\vec{r})$ and $S_{0}(\vec{q})$ is the
structure factor of a Gaussian chain without interaction:

\begin{equation}
S_{0}(\vec{}q) = N V \rho_0\ f((qR_{G})^2), \label{S0eq}
\end{equation}

In Eq.\ref{S0eq}, $N$ is the number of monomers per chain, $R_G$
the radius of gyration of one chain, $V$ the total volume,
$\rho_0$ the homogeneous equilibrium density of monomers and
$\displaystyle{f(x) = \frac{2}{x^2}\ (e^{-x} + x - 1) }$ the Debye
function.

Hence, Eq.\ref{RPAeq} allows one to calculate the structure factor
of semi-dilute solutions of polymers for any given interaction
potential between monomers. This equation can be used for hairy
wormlike micelles, once a phenomenological potential
$\mathbf{V}(\vec{r})$ is given to account for the steric
repulsion.

\subsection{Phenomenological repulsive potential}

We assume that the interaction potential between hairy polymers
can be considered as the sum of a standard excluded volume
potential (polymer/solvent interactions), $v_{0}\delta(\vec{r})$,
and an additional repulsive potential, $V_g(\vec{r})$, due to the
steric layer. The two physical criteria for this steric potential
are that $V_g(\vec{r})$ should be soft and short range. We thus
choose to model it with a Gaussian function, which decreases
sufficiently fast to be considered as short range:

\begin{equation}
V_g(\vec{r}) = \ U_0\ \rm{exp}(-\frac{r^2}{2\delta^2})
\end{equation}

We expect $\delta$, the width of the Gaussian, to be on the order
of the steric layer thickness. We moreover expect both $\delta$
and the amplitude of the repulsive potential, $U_0$, to increase
as the amount of copolymer $\alpha$ increases. We note that a
Gaussian form for the potential has been recently justified for
some soft interacting objects, such as polymer coils
\cite{louis2000,bolhuis2001}, flexible dendrimers
 \cite{likos2002}, or star polymers near the $\theta$-point
\cite{graf1998}. A Gaussian shape has the advantage of leading to
a simple analytical expression for the structure factor of hairy
polymers:

\begin{equation}
V S^{-1}(\vec{q}) = V S_{0}^{-1}(\vec{q}) +  v_{0} + U_0\
(2\pi\delta^2)^{3/2}\ \rm{exp}(-\frac{(q \delta)^2}{2})
\label{FS1}
\end{equation}

The last two terms of the right end side of Eq.\ref{FS1} represent
the interaction part of the structure factor: $v_{0}$ is the
excluded volume parameter and the last term is the Fourier
transform of $V_g(\vec{r})$.


\subsection{Theoretical structure factor and comparison with experiments}

We take advantage of the close analogy between classical polymers
and giant micelles and apply the RPA results to semi-dilute
solutions of wormlike micelles. We use Eq.\ref{FS1} to compute the
structure factor of solutions of hairy micelles and investigate
their variation with the micellar concentration and
characteristics of the copolymer-induced Gaussian potential.

We define the statistical unit or ``monomer'' as a slice of
micelle of length $a$ equal to $2\ l_{P}$, where $l_{P}\simeq190
\, \rm{\AA}$ is the persistence length of the micelle, and of
radius $r_{0}\simeq30 \, \rm{\AA}$
\cite{appell-bassereau1989,appell-marignan1991}. The equilibrium
density of monomers, $\rho_0$, is then related to the surfactant
volume fraction by $\rho_0 = \phi / 2 \pi r_0^2 l_P$. The number
of statistical units per chain obeys $\displaystyle{N =
\frac{a_0^2}{4 \pi r_0 l_P}\
\left(\frac{v_{sol.}}{v_{t.a.}}\right)^{-1/2}\ \phi^{1/2}\
exp[E/2k_{B}T] }$ \cite{cates-candau1990}, where $a_0 =7.2 \,
\rm{\AA}$ is the surfactant polar head diameter, $E\approx 26 \,
\rm{k_{B}T}$ is the end-cap energy  and $v_{sol.}\approx 30 \,
\rm{\AA}^{3}$ and $v_{t.a.}\approx 595 \, \rm{\AA}^{3}$ are the
volume of a molecule of solvent and of surfactant respectively. We
use this expression for $N$ to calculate the radius of gyration of
a chain: $R_G=a\sqrt{N}/\sqrt{6}$. The excluded parameter is fixed
to $v_0 = a^3$, which corresponds to a polymer in an athermal
solvent.

With excluded volume as unique interaction potential, Eq.\ref{FS1}
is reduced to the classical expression first derived by Edwards
\cite{doi-edwards} for a polymer chain with excluded volume
interactions and the structure factor is a monotonically
decreasing function of the $q$ vector. By contrast, in the
presence of a Gaussian potential, we show that the structure
factor $S(q)$ may have a non-monotonic variation and exhibit a
peak at a finite wavevector $q^{*}$. The condition for the
existence of a peak can be determined from the analytical
expression of the structure factor (Eq.\ref{FS1}). This is easily
calculated if we note that the derivative of the inverse of the
Debye function, $f^{-1}(x)$, tends to $1/2$ for large $x$, while
it tends to $1/3$ when $x$ is small. Assuming that $(q R_G)^2 \gg
1$, which is always verified in the neighbourhood of the peak, we
obtain that the structure factor displays a peak for surfactant
volume fraction $\phi$ larger than a threshold value $\phi_{c}$:

\begin{equation}
\phi_{c} = \frac{4 \pi r_{0}^2 l_{P}^3}{3\ (2 \pi)^{3/2}}\
\frac{1}{\delta^5 U_0} \label{phiC}
\end{equation}

\noindent and the position $q^{*}$ of the peak is given by:

\begin{equation}
q^* = \frac{\sqrt{2}}{\delta} \ \sqrt{ln \left(\frac{\phi}{\phi_c}
\right)}=
  \frac{\sqrt{2}}{\delta} \ \sqrt{ln \left(\frac{3\ (2
\pi)^{3/2}}{4 \pi r_{0}^2 l_{P}^3}\ \phi\ \delta^5 U_0 \right)}
\label{qmax}
\end{equation}

The critical volume fraction depends on the two parameters
characterizing the Gaussian potential, $\delta$ and $U_{0}$. It
decreases as either the range or the amplitude of the potential
increases, but $\phi_{c}$ is more sensitive to $\delta$ than to
$U_0$. Note that $\phi_{c}$ diverges if the unique repulsive
potential is the excluded volume ($\delta=0$ or $U_{0}=0$),
consistently with the structure factor being strictly decreasing,
as mentioned above.

\begin{figure}
\begin{center}
\includegraphics[width=0.5\textwidth]{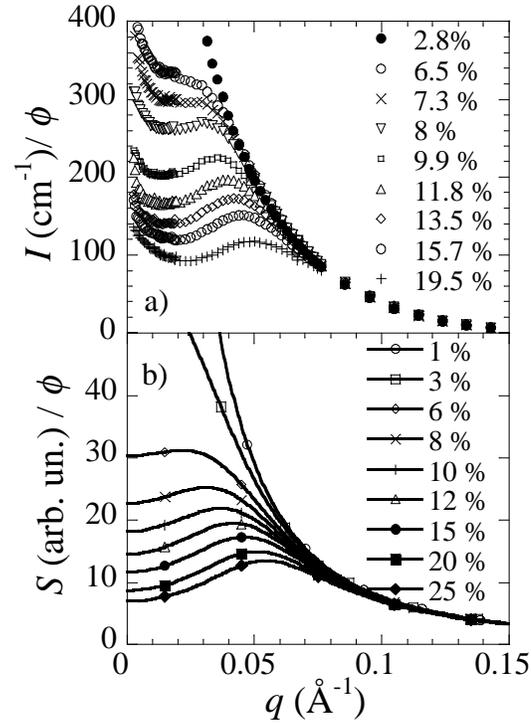}
\end{center}
\caption{(a) Experimental scattering profiles and (b) theoretical
structure factors, normalized by surfactant volume fraction
$\phi$. Curves are labeled by surfactant volume fraction $\phi$.
In (a), the copolymer (F108) over surfactant ratio is $\alpha=1\
\%$; in (b), the amplitude and range of the Gaussian potential are
$U_{0}=800 \ k_{B}T$ and $\delta= 34 \ \rm{\AA}$ respectively.
}\label{exp-phivar}
\end{figure}

\begin{figure}
\begin{center}
\includegraphics[width=0.5\textwidth]{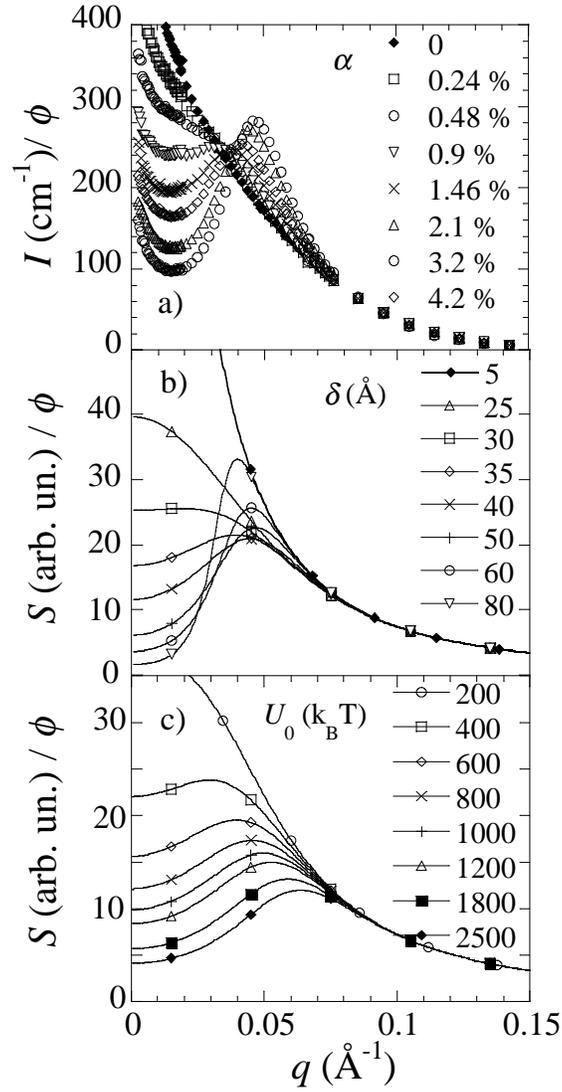}
\end{center}
\caption{(a) Experimental scattering profiles and (b, c)
theoretical structure factors, normalized by surfactant volume
fraction $\phi$. In (a), the surfactant volume fraction is
$\phi=9\ \%$ and curves are labeled by copolymer (F108) over
surfactant ratio $\alpha$. In (b) and (c), the surfactant volume
fraction is $\phi=10\ \%$. In (b) the amplitude of the potential
is $U_{0}=800\  k_{B}T$ and curves are labeled by range of the
potential $\delta$.  In (c), the range of the potential is
$\delta= 34\ \rm{\AA}$ and curves are labeled by amplitude of the
potential $U_0$.}\label{exp-alphavar}
\end{figure}

In order to  directly compare the theoretical structure factors
with experiments, we first choose fixed values for the parameters
of the Gaussian potential, which should correspond to keeping the
copolymer over surfactant molar ratio $\alpha$ constant, and we
vary the surfactant volume fraction. The theoretical structure
factors obtained for $\delta = 34\ \, \rm{\AA}$ and $U_0 = 800 \,
\rm{k_BT}$ are plotted Fig.\ref{exp-phivar}b and exhibit features
very similar to the experimental scattering profiles
(Fig.\ref{exp-phivar}a). At low $\phi$, the structure factor is a
decreasing function of the wavevector. By contrast, for $\phi
\simeq 6\ \%$, a correlation peak appears, which becomes more and
more narrow and whose position is reported to higher wavevector as
$\phi$ increases. On the other hand, when the copolymer over
surfactant molar ratio $\alpha$ is experimentally varied, the
copolymer-induced steric potential changes and thus the two
characteristic parameters of the Gaussian potential should change
as well. However, there is no clear intuitive argument how to
determine the influence of $\alpha$ on $\delta$ and $U_0$
separately. To mimic experimental data taken at various $\alpha$,
we therefore vary independently $\delta$ and $U_0$. The structure
factors obtained for $\phi=10 \, \%$, $U_0=800\, \rm{k_{B}T}$ and
  $\delta$ in the range $5-80 \, \rm{\AA}$ are plotted in
Fig.\ref{exp-alphavar}b, while the structure factors
  obtained for $\phi=10 \, \%$, $\delta=30\, \rm{\AA}$ and
$U_0$ in the range $200-2500\, \rm{k_{B}T}$ are plotted in
Fig.\ref{exp-alphavar}c. Both series display features very similar
to the experimental scattering profiles shown in
Fig.\ref{exp-alphavar}a. For a fixed amplitude $U_0$ of the
potential, a correlation peak appears for $\delta$ larger than
$30\ \rm{\AA}$ and becomes more pronounced as $\delta$ increases.
Similarly, increasing $U_0$ with fixed $\delta$ leads to the
emergence of a peak and to an increase of its intensity. In the
two cases, similarly to what is obtained experimentally upon
increasing $\alpha$, the parameters $\delta$ and $U_0$ have poor
influence on the peak position.

Thus, the theoretical structure factors capture the essential
features of the experimental scattering profiles and their
evolution with the experimental parameters. However, the
comparison between theory and experiment can only be qualitative
since theory describes the correlation between the objects and
does not take into account the form of the scattering objects,
whereas the scattered intensity is experimentally measured. While
the relation between scattered intensity, $I(q)$, and structure
  and form  factors, $S(q)$ and $P(q)$, is simple for spherical
objects ($I(q) = P(q) S(q)$), it is more complex for semi-dilute
solutions of linear and flexible objects. Moreover, the RPA
technique usually does not describe correctly the structure factor
at very low $q$. These two limitations make it more difficult to
fit precisely the experimental scattering profiles, though
remarkable agreement with the experimental position of the peak
can be found, as shown in the next section.

\section{Discussion}
\label{SEC:discuss}

\subsection{Fit of the peak position}

\begin{figure}
\begin{center}
\includegraphics[width=0.5\textwidth]{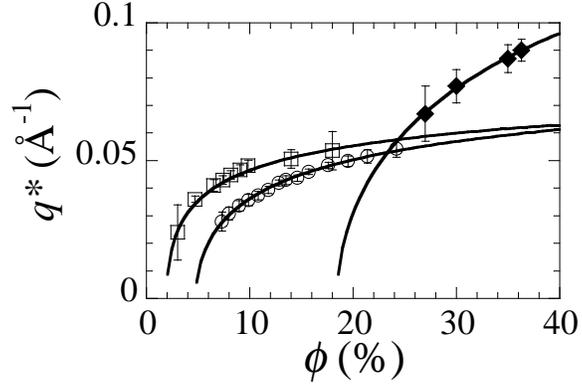}
\end{center}
\caption{Variation of the peak position with surfactant volume
fraction for samples without copolymer (diamonds), and with a
copolymer (F108) over surfactant ratio $\alpha = 1\ \%$ (empty
circles) and $\alpha = 3.2\ \%$  (empty squares). Symbols are
experimental data points and lines are best fits using
Eq.\ref{qmax}. The fitting parameters are given in table
\ref{table1}. }\label{fitqmax1}
\end{figure}

To analyze more quantitatively our data, we assume that the
experimental peak position in the scattered intensity is correctly
described by the theoretical structure factor. For surfactant
volume fraction above the critical surfactant volume fraction
$\phi_{c}$, the peak position $q^{*}$ is given by Eq.\ref{qmax}.
We use this equation to fit the experimental $\phi$ dependence of
$q^{*}$, with $\delta$ and $U_{0}$ as fitting parameters. As shown
in fig.\ref{fitqmax1}, a very good fit is obtained for the
experimental data obtained for hairy micelles decorated with
different amounts of copolymer F108 as well as for naked micelles.
An equally good agreement is obtained for the two other copolymers
(data not shown). The results of the fits are given in Table
\ref{table1} for all experimental configurations.

In the case of naked micelles, a correlation peak is detected only
at  a very large concentration ($\phi$ above $24\, \%$) and the
$\phi$ dependence of $q^{*}$ can hence be fitted only in a reduced
range of concentrations. We find for the fitting parameters $U_0=
24604\ \rm{k_{B}T}$ and $\delta = 13\ \rm{\AA}$, which correspond
to a very high and narrow potential. For hairy micelles, the fits
always extend over a larger interval of concentrations than that
for naked micelles. The values of $U_0$ range between $790$ and
$2770 \, \rm{k_BT}$ and those of $\delta$ range between $26$ and
$39\ \rm{\AA}$. Hence, the range of the potential is on the order
of the radius of gyration of the copolymer, and is always larger
than that of naked micelles. It moreover increases from $33.7$ to
$38.7\ $\AA\ when $\alpha$ increases from $1$ to $3.2\ \%$ (for
F108 copolymer). In addition, we find $\delta$ smaller for the
copolymer F68 than for the copolymers PC18 or F108, as expected,
since F68 possesses shorter hydrophilic chains than F108 or PC18.
On the other hand, we find that, in the presence of the copolymer
layer, the amplitude of the potential is considerably reduced
compared to the case of naked micelles. This can be intuitively
understood. Indeed, the copolymer layer covering the micelles is
presumably very compressible, since this layer is highly swollen
by the solvent (the regime of a dry brush is never reached
experimentally, $\alpha$ being always comparable to the overlap
threshold $\alpha^{*}$). This should result in a small value for
$U_{0}$, much smaller than for naked micelles for which the dense
shell of surfactant polar heads is very little compressible. In
fact, for naked micelles, a potential close to a hard core
potential is expected, consistently with our results. Although the
values of the amplitude $U_{0}$ that we extract from the fits may
seem very high, they correspond to very reasonable values for the
mean free energy per anchored polymer tail of roughly 15 $k_{B}T$.

One could \textit{a priori} separate the contribution of the
surfactant shell from that of the copolymer layer, by replacing
the Gaussian potential in Eq.\ref{RPAeq} by the sum of two
Gaussian functions, one accounting for the surfactant contribution
and the other one for the copolymer contribution. The former
Gaussian function is expected to be very narrow and high, while
the latter is expected wider and lower. We can compute the
structure factors in this approach, by taking for the former
potential the parameters derived from the fit of the naked
micelles and letting free the parameters ($\alpha^{\rm{copo}}$ and
$U_{0}^{\rm{copo}}$) for the copolymer contribution. In this case,
we find that the essential features of the theoretical structure
factors and their evolution with $\phi$, $\alpha^{\rm{copo}}$ and
$U_{0}^{\rm{copo}}$ remain unchanged. Moreover, the values of
$\alpha^{\rm{copo}}$ and $U_{0}^{\rm{copo}}$, derived from the
$\phi$ dependence of $q^{*}$, which should account solely for the
copolymer layer, are of the same order of magnitude as the values
obtained with one Gaussian function, although $U_{0}^{\rm{copo}}$
is slightly smaller than $U_{0}$. This strongly suggests that only
the tail of the potential (of energy at most of a few $k_{B}T$) is
important and that this tail does not vary much with the addition
of a high and narrow potential. Moreover it is clear that the
concavity of the potential $V_g(\vec{r})$ is non-physical for $r <
\delta$, limiting the validity of a Gaussian potential to not too
short distances.

Finally, we compare the theoretical values of $\phi_c$, deduced
from the fitting parameters using Eq.\ref{phiC}, to the
experimental concentrations. As can be seen in Table \ref{table1},
the experimental values show the same variations as the
theoretical ones but are systematically smaller. This discrepancy
could originate from the fact that what is experimentally measured
is the scattered intensity $I$. Peaks of low magnitude in the
structure factor may thus be masked in a $I$
 \textit{vs} $q$ plot because of the form factor of the
objects, which is a decreasing function of the $q$ vector.

\subsection{Effective thickness of the copolymer layer}

The pair of fitting parameters, $\delta$ and $U_0$, allows a
determination of an effective thickness of the copolymer layer,
$h$. The simple physical criterion we apply is based on the
assumption that  the micelles enter in contact as soon as their
interaction potential overcomes the thermal energy $k_{B}T$. Thus,
for a monomer-monomer distance $r=2(r_{0}+h)$, the Gaussian
potential is equal to $1\, k_{B}T$ and
$V_{g}(r)=\displaystyle{U_0\ exp\ (-r^2/2\ \delta^2) = 1}$. This
criterion leads to a relation between the effective thickness $h$,
the naked radius of the micelles $r_0$, and the potential
parameters $\delta$ and $U_0$:
\begin{equation}
r_{0}+h = \delta \  \sqrt{(\ln U_0) / 2}
\end{equation}

Using this simple criterion for naked micelles, we obtain
$r_{0}=29.2 \, \rm{\AA}$ ($h$ being equal to $0$ by definition in
this case), a value in striking good agreement with the expected
value ($r_{0}\simeq30 \, \rm{\AA}$
\cite{appell-bassereau1989,appell-marignan1991}). The values of
$h$ deduced from the fitting parameters $\delta$ and $U_0$ are
reported in table \ref{table1} for hairy micelles and are very
close to the radii of gyration of the copolymers. They are also
very close, although slightly larger, to the values deduced from
two other independent methods \cite{MassieraLiqMatt}. They
moreover follow the expected trends: $h$ increases with the amount
of copolymer and is larger when the polymer is longer.

\begin{table}[h]
\centering
\begin{tabular}{|c||c|c|c|c|c|}
\hline & $\delta$ (\AA) & $U_0$ ($k_{B}T$) & $h$(\AA)& theoretical
$\phi_c$ & experimental $\phi_c$\\ \hline\hline

$\alpha = 0$ & $13$ & $24604$ &  & $18\ \%$ & $25.5\pm 1.5\ \%$\\
\hline

F108, $\alpha = 1\ \%$ & $33.7$ & $793$& $31.6$ & $4.8\ \%$ &
$7.3\pm 0.7\ \%$\\ \hline

F108, $\alpha = 3.2\ \%$ & $38.7$ & $1007$ & $41.9$ & $1.9\ \%$ &
$3\pm 1\ \%$\\ \hline

F68, $\alpha = 2.1\ \%$ & $25.7$ & $2768$ & $21.2$ & $5.3\ \%$ &
$8\pm 1\ \%$\\ \hline

PC18, $\alpha = 2.1\ \%$ & $33.6$ & $1196$ & $33.2$ & $3.2\ \%$ &
$4.9\pm 1.9\ \%$\\ \hline
\end{tabular}
\caption{Fitting parameters, range $\delta$ and amplitude $U_0$ of
the Gaussian potential, effective thickness of the copolymer layer
$h$ and theoretical and experimental critical volume fraction
above which a correlation peak appears, for samples with different
copolymers and different amounts of copolymer.} \label{table1}
    \end{table}

\subsection{Microscopic model for the copolymer-induced repulsive potential}

At this point of the discussion, one can finally raise the
question of the microscopic origin of the Gaussian potential used
here. With this function for the potential,
  the model
describes correctly the behavior of the scattering profiles and
especially  the  variations of the peak position with $\phi$. It
appears nevertheless difficult to draw a precise link between the
microscopic details of the system and the effective mean-field
potential. Qualitatively, one can however suggest the following
physical picture: the Gaussian potential originates from the
copolymer layer covering the wormlike micelles, inducing thereby
an additional steric repulsion. Two regimes have to be considered.
In the brush regime the micelles are covered with a semi-dilute
copolymer layer, while in the mushroom regime the copolymer chains
are isolated on the micelles. The interaction is clearly stronger
and with a larger range in the brush regime. In order to bring two
micelles close to each other, a large energy is needed to
compensate the energy cost to compress the copolymer layer below
its equilibrium thickness value. This compression energy leads to
a strong repulsion which is relevant for short distances $r$
between micelles. This energy can be evaluated from the gap in
energy between the equilibrium brush free energy and its value at
$r$ in the brush regime, while in the mushroom regime the energy
can be evaluated from the energy cost for confining the polymer on
distances smaller than its Flory radius. Although the theoretical
potentials calculated with this approach \cite{thesis-massiera}
cannot be satisfyingly fit with a Gaussian function, they give
numerical values of $h$ in excellent agreement with those deduced
from the $q^*$ \textit{vs} $\phi$ fits.

\section{CONCLUSION}

\label{SEC:concl}

To conclude, we have shown that a RPA approach starting from an
Edwards Hamiltonian with a soft Gaussian repulsive potential
reproduces well the experimental results obtained for the
structure of hairy wormlike micelles, and in particular the
variation of the correlation peak with the micellar concentration
$\phi$.
  This model
allows one to extract physical parameters with very reasonable
numerical values.

One can ask whether the RPA is physically justified in the type of
systems studied here. Because it essentially neglects strong
density fluctuations (see Appendix and
\cite{doi-edwards,DeGennes-scalingconcepts}), this approximation
is \textit{a priori} best suited for concentrated solutions or
melts of polymer.
  However, this model can
nevertheless be applied to less concentrated solutions, provided
fluctuations are weak \cite{doi-edwards}. In
Ref.\cite{muller2000}, the validity of RPA is discussed in the
case of semi-dilute solutions. The authors show that a
renormalization of the excluded volume parameter leads to a very
good agreement between the RPA and the renormalization group
theory. Furthermore, in the case of giant micelles, one can expect
to reach a concentrated regime even at relatively low surfactant
concentration because of the large  persistence length of the
micelles. A concentrated regime is indeed attained as soon as the
correlation length $\xi$ is on the order of $l_P$, which
eventually occurs in the range of concentration investigated
experimentally. However, in the same range of concentration,
several rheology \cite{massiera2002} and light scattering
\cite{berret-porte1993} experiments indicate that $\xi$ varies as
 $\phi^{-3/4}$, a scaling characteristic of a
semi-dilute solution, in seeming contradiction with the system
being in a concentrated regime. This apparent discrepancy
underlines how the  border between semi-dilute and concentrated
solutions is ill defined for wormlike micelles solutions.

Finally, we believe that hairy wormlike micelles solutions provide
an original experimental system to illustrate the influence of a
soft short range repulsion in isotropic solutions of linear
flexible objects. The interaction induced by the copolymer layer
is soft but sufficiently strong to induce a correlation peak in
the structure factor. The Random Phase Approximation has proven to
be helpful in describing the qualitative behavior of the scattered
patterns. Such a model may be used in a variety of ``hairy linear
objects'' such as copolymer micelles, or hairy polymers (comblike
polymers), for which the interaction is short range.


\clearpage

\appendix*
\section{}
In this appendix, we detail the structure factor calculation for
the Random Phase Approximation (RPA).

From the hamiltonian in $k_BT$ units (Eq.\ref{free-energy-RPA}),
\begin{equation}
\displaystyle{{\LARGE \mathcal{H}}({\vec{r}(s)}) = \frac{3}{2a^2}\
\int_{0}^{N} ds \left( \pfrac{\vec{r}}{s} \right)^2 +\frac{1}{2}
\int \int_{0}^{N} ds\ ds'\ \mathbf{V}\left(\vec{r}(s) -
\vec{r}(s')\right)},
\end{equation}
we can evaluate the chain partition function ${\LARGE
\mathcal{Z}}$ as the sum, over all the configurations
$\vec{r}(s)$, of the Boltzmann factor:
\begin{equation}
{\LARGE \mathcal{Z}} = \displaystyle{\int \mathfrak{D}\vec{r}(s) \
\rm{exp} \left[- \mathcal{H}(\vec{r}(s)) \right]}
\end{equation}

\noindent The local monomer density is defined as:
\begin{equation}
\rho (\vec{r}(s)) = \int_{0}^{N} ds\ \delta\left(\vec{r} -
\vec{r}(s)\right).
\end{equation}

\noindent
From the definition of $\rho (\vec{r}(s))$, we can write
the identity:
\begin{equation}
\int \mathfrak{D}\rho(\vec{r})\ \delta \left(\rho (\vec{r}(s)) -
\int_{0}^{N} ds\ {\LARGE \delta}\left(\vec{r} - \vec{r}(s)\right)
\right) = 1, \label{rho1}
\end{equation}
which can also be expressed in the Fourier functional space:
\begin{equation}
\int \mathfrak{D}\rho(\vec{r}) \int
\mathfrak{D}\widehat{\rho}(\vec{r})\ \rm{exp} \left[i \int
\it{d\vec{r}}\ \widehat{\rho}(\vec{r}) \rho(\vec{r})- i \int
\it{ds} \widehat{\rho}(\vec{r}(s)) \right] = 1, \label{rho2}
\end{equation}
By introducing this identity into the partition function, ${\LARGE
\mathcal{Z}}$ can be expressed as a function of collective
variables ${\rho(\vec{r})}$ and ${\widehat{\rho}(\vec{r})}$, the
local density and its conjugate variable in  Fourier space:
\begin{eqnarray}
{\LARGE \mathcal{Z}} &=& \int \mathfrak{D}\rho(\vec{r})\
\mathfrak{D}\wrho(\vec{r})\ \rm{exp} \left[
-\mathcal{F}(\{\rho(\vec{r}),\widehat{\rho}(\vec{r})\}) \right],\
\label{Z1}
\\
\rm{with} \ \ {\LARGE \mathcal{F}}
(\{\rho(\vec{r}),\widehat{\rho}(\vec{r})\}) &=& -\rm{ln} \big(
\zeta (\{i \wrho(\vec{r})\}) \big) -i\ \int d\vec{r} \
\wrho(\vec{r})\ \rho(\vec{r}) + \frac{1}{2} \int d\vec{r}\ d\vrp
\rho(\vec{r}) \mathbf{V}(\vec{r}-\vrp)\ \rho(\vrp) \nonumber
\\ \label{F1}
\end{eqnarray}
\noindent and where $\zeta (\{i \wrho(\vec{r})\})$ is a  function
of $\wrho(\vec{r})$ only:
\begin{equation}
\zeta (\{i \wrho(\vec{r})\}) = \int \mathfrak{D} \vec{r}(s)\
\rm{exp} \left[- \frac{3}{2a^2}\ \int ds
\left(\pfrac{\vec{r}(s)}{s}\right)^2\ \right]\ \rm{exp} \left[- i\
\int d\vec{r}\ \widehat{\rho}(\vec{r})\ \int ds\ \delta(\vec{r} -
\vec{r}(s)) \right]
\end{equation}

By minimizing the free energy ${\LARGE \mathcal{F}}
(\{\rho(\vec{r}),\widehat{\rho}(\vec{r})\})$ with respect to both,
$\rho(\vec{r})$ and $\wrho(\vec{r})$, we obtain $\rho_{0}$ and
$\wrho_{0}$, the equilibrium homogeneous densities:
\begin{displaymath}
\left\{ \begin{array}{ll} \displaystyle{ \left.
\pfrac{\mathcal{F}}{\rho(\vec{r})} \right|_{\rho_{0}} = 0 } \\ \\
\displaystyle{\left. \pfrac{\mathcal{F}}{i \wrho(\vec{r})}
\right|_{i \wrho_{0}} = 0}\end{array} \right. \qquad
\Longleftrightarrow \qquad \left\{
\begin{array}{ll} \displaystyle{\rho_{0} = \frac{N}{V}}  \\ \\
\displaystyle{i\wrho_{0} = \rho_{0} W}\end{array}\right.
\end{displaymath}
where $\displaystyle{W = \int d\vec{r}\ \mathbf{V}(\vec{r})}$ and
$V$ is the total volume.

Assuming fluctuations are weak, we develop a perturbative calculus
around the equilibrium homogeneous densities
$\{\rho_{0},\widehat{\rho}_{0} \}$:
\begin{displaymath}
\left\{ \begin{array}{ll} \displaystyle{\rho(\vec{r}) = \rho_{0} +
\delta\rho(\vec{r})}
\\ \displaystyle{\wrho(\vec{r}) = \wrho_{0} +
\delta\wrho(\vec{r}),}\end{array} \right.
\end{displaymath}

In ${\LARGE \mathcal{Z}}$, we only keep constant terms (which
contribute to the prefactor ${\LARGE \mathcal{Z}}_{0}$) , and the
terms of second order in
$\{\delta\rho(\vec{r}),\delta\widehat{\rho}(\vec{r}) \}$, the sum
of the first order terms being equal to zero. The partition
function thus reads:
\begin{eqnarray}
{\LARGE \mathcal{Z}} &=& {\LARGE \mathcal{Z}}_{0} \ \int
\mathfrak{D}\delta\rho(\vec{r})\ \mathfrak{D}\delta\wrho(\vec{r})\
\rm{exp} \left[ i \int d\vec{r}\ \delta\rho(\vec{r})\
\delta\wrho(\vec{r}) \right] \ \rm{exp} \left[ -\frac{1}{2} \int
d\vec{r}\ d\vrp\ \delta\rho(\vec{r})\ \mathbf{V}(\vec{r}-\vrp)\
\delta\rho(\vrp) \right]{} \nonumber \\ & &{} \rm{exp} \left[
-\frac{1}{2} \int d\vec{r}\ d\vrp\ \delta\wrho(\vec{r})\
\frac{\rho_{0}}{V} \ \it{g_{D}} (\vec{r}-\vrp) \
\delta\widehat{\rho}(\vrp) \right] \label{Zr}
\end{eqnarray}
The function $g_{D} (\vec{r}-\vrp)$ is the correlation function of
a Gaussian chain:
\begin{equation} g_{D} (\vec{r}-\vrp) = \frac{V^2}{N}\ \int
\mathfrak{D}\vec{r}(s) \rm{exp}\left[- \frac{3}{2a^2}\ \int ds
\left(\pfrac{\vec{r}(s)}{s}\right)^2\right]\ \int ds\
\delta(\vec{r}-\vec{r}(s))\ \int ds'\ \delta(\vec{r}'-\vec{r}(s'))
\end{equation}
Finally, in Fourier space, the partition function reads:
\begin{displaymath}
{\LARGE \mathcal{Z}} = {\LARGE \mathcal{Z}}_{0} \ \int
\mathfrak{D}\delta\rho(\vq)\ \mathfrak{D}\delta\wrho(\vq)\
\rm{exp} \left[ -\frac{1}{2} \int d\vq\
\big(\delta\rho(\vq),\delta\wrho(\vq) \big)\ \mathbb{A}(\vq)\
{\delta\rho(-\vq) \choose \delta\wrho(-\vq)} \right]
\end{displaymath} where the matrix $\mathbb{A}(\vq)$ is equal to:
\begin{displaymath}
\mathbb{A}(\vq) = V^2 \ \left( \begin{array}{cc} V
\widetilde{\mathbf{V}}(\vq) & -i
\\ -i & \rho_{0} g_{D}(\vq),
\end{array} \right)
\end{displaymath}
where $\widetilde{\mathbf{V}}(\vq)$ is the Fourier transform of
the interaction potential:
\begin{equation}
\widetilde{\mathbf{V}}(\vq) = \frac{1}{V}\ \int d\vec{r}
e^{-\vq.\vec{r}} \mathbf{V}(\vec{r})
\end{equation}
Finally, the Fourier transform of the density fluctuation
correlation, which is proportionnal to the structure factor,
reads:

\begin{equation}
S(\vq) = V^2 \langle \delta\rho(\vq)\ \delta\rho(-\vq) \rangle =
V^3\ \left[ \mathbb{A}^{-1}(\vq) \right]_{11}.
\end{equation}

We thus obtain the simple form of Eq.\ref{RPAeq} for the structure
factor $S(\vq)$:
\begin{equation}
S^{-1}(\vq)  =  S_{0}^{-1}(\vq) + \ \widetilde{\mathbf{V}}(\vq)
\end{equation}
where $S_{0}(\vq) = N \rho_0 V\ f((q R_G)^2)$, and
$\displaystyle{f(x) = \frac{2}{x^2} (e^{-x}+x-1)}$ is the Debye
function.

\begin{acknowledgments}
We are grateful to R. Aznar for the synthesis of the PC18. Local
contacts, L. Auvray at LLB and B. Dem\'{e} and J. Zipfel at ILL,
are acknowledged. E.P thanks Henri Orland for very useful
discussions.
\end{acknowledgments}


\end{document}